# OVERCOMING OBSTACLES TO INTEGRABILITY
# IN PERTURBED NONLINEAR EVOLUTION EQUATIONS

Alex Veksler[1] and Yair Zarmi[1,2]
Ben-Gurion University of the Negev, Israel
[1]Department of Physics, Beer-Sheva, 84105
[2]Department of Solar Energy & Environmental Physics
Jacob Blaustein Institute for Desert Research, Sede-Boqer Campus, 84990

ABSTRACT

Obstacles to integrability in perturbed evolution equations are overcome by allowing higher-order terms in the expansion of the solution to depend explicitly on time and position. With a special expansion algorithm, obstacles vanish explicitly for single-waves zero-order approximations, and exponentially, outside the wave interaction-region, for multiple-waves. Simple expressions for their asymptotic effects on the solution are obtained.

Corresponding author
Yair Zarmi
Jacob Blaustein Institute for Desert Research
Sede-Boqer Campus, 84990, Israel
Tel: (972)-8-659-6920      Fax: (972)-8-659-6921
e-mail: zarmi@bgumail.bgu.ac.il

There are quite a few integrable nonlinear evolution equations the solutions of which are of physical interest [1], e.g., fronts (Burgers equation), and single or multiple solitons (KdV, NLS and other equations). When a small perturbation is added to such equations, they are often analyzed by the method of Normal Forms (NF) [2-4]. The motivation is the expectation that the dynamical equation for the zero-order approximation to the solution (the NF) is also integrable and preserves the nature of the unperturbed solution. However, the analysis often leads to the emergence of *obstacles to integrability* [5-12]. These are terms that the perturbative expansion of the dynamical equation generates, which cannot be accounted for by the formalism. Obstacles to integrability do not appear when the zero-order solution is a single wave, e.g. a single front or a single soliton; the NF then amounts to no more than updating of the dispersion relation that determines the wave velocity [2-4]. In the general case (e.g., multiple-wave solutions), obstacles do emerge, except for specific forms of the perturbation [8-12]. To simplify the construction of the perturbative expansion of the solution (Near-Identity Transformation - NIT), the usual practice has been to include these unaccountable terms in the NF. This makes the NF nonintegrable, hence the name "obstacles to integrability". Moreover, inclusion of obstacles in the NF may spoil features of physical interest of the solution for the zero-order term (e.g., its wave structure).

These difficulties are consequences of the assumption, usually made in the NF expansion, that all the terms in the NIT are differential polynomials in the zero-order approximation and do not depend explicitly on the independent variables ($t$ and $x$). The problem arises already in the NF analysis of ODE's if it is assumed that higher-order corrections in the NIT do not depend explicitly on time. This assumption works for systems that are described by autonomous equations with a linear unperturbed part. Inconsistencies may emerge in other cases, unless explicit dependence on time is allowed.

We show that by giving up this assumption, obstacles can be accounted for through the solution of the (homological) equation that determines the NIT. They cease to be obstacles to integrability or to the application of the method of NF. The NF remains integrable; its solution (the zero-order approximation) retains the character of the unperturbed solution. In some cases, the asymptotic behavior of the solution of the NIT may be expressed in closed form. With a judicious choice of the structure of the higher-order terms in the NIT, the obstacles assume a "canonical" form that manifestly ensures their vanishing in the case of a single-wave solution. Focusing on examples of two-wave solutions, it is found that the effect of the obstacles in the canonical form is confined to the region within which interaction among the waves is significant in the zero-order approximation. The details of this brief report will be published elsewhere.

Consider a perturbed autonomous PDE of the general form

$$w_t = S_2[w] + \sum_k \varepsilon^k F_k[w] \tag{1}$$

(Square brackets imply that the corresponding term is a differential polynomial in *w*. The notation for the unperturbed part in Eq. (1) will become clear in the following.) The unperturbed equation

$$\tilde{w}_t = S_2[\tilde{w}] \tag{2}$$

is assumed to be integrable. ε is a physically or calculationally justified small parameter. We assume a power-series expansion for *w* (the *Near Identity Transformation* – NIT):

$$w = u + \varepsilon u^{(1)} + \varepsilon^2 u^{(2)} + \ldots \tag{3}$$

The evolution of the zero-order term, *u*(*x*, *t*), is governed by the *Normal Form* (NF):

$$u_t = S_2[u] + \varepsilon a_1 S_3[u] + \varepsilon^2 a_2 S_4[u] + \ldots \tag{4}$$

In Eq. (4), the coefficients $a_n$ are determined by the structure of the original equation, Eq. (1). $S_n[u]$, $n \geq 3$, are *symmetries* of the unperturbed term, $S_2[u]$, generated by the requirement that their Lie brackets with $S_2$ vanish [13]:

$$[S_2[u], S_n[u]] \equiv \sum_i \left\{ \frac{\partial S_2}{\partial u_i} \partial_x^i S_n - \frac{\partial S_n}{\partial u_i} \partial_x^i S_2 \right\} = 0 \qquad (u_i \equiv \partial_x^i u) \qquad (5)$$

The Lie brackets of any two symmetries also vanish.

While the structure of all members $S_n$, $n \geq 3$, of the *hierarchy* of symmetries is determined by $S_2$, in hierarchies of many evolution PDE's, the lowest member, $S_1$, is

$$S_1 = u_x \qquad (6)$$

For later use, we mention that $S_n[u]$ can be written as gradients of other differential polynomials:

$$S_n[u] = \partial_x G_n[u] \qquad (7)$$

The correction term, $u^{(n)}$, is determined by the $n$'th-order *homological equation*, which is obtained by substitution of Eqs. (3) and (4) in Eq. (1) and expansion of the result in powers of $\varepsilon$:

$$a_n S_{n+2}[u] + \partial_t u^{(n)} = [S_2[u], u^{(n)}] + Z_n[u] \qquad (8)$$

The derivative with respect to time on the l.h.s. of Eq. (8) applies only to the *explicit* dependence of $u^{(n)}$ on $t$, if such dependence exists. $Z_n[u]$ is the collection of all the known terms that emerge in this order, including the contribution of the perturbation as well as of $u^{(k)}$, $k < n$, which have been solved for in previous orders. (For instance, in the $O(\varepsilon)$ equation, $Z_1[u] = F_1[u]$.)

Consider, first, the case of a *single-wave solution* of the NF: $u(x,t) = u(\xi)$ ($\xi = x - vt$). Assuming that the wave velocity itself may be expanded in a power series:

$$v = v_0 + \varepsilon v_1 + \varepsilon^2 v_2 + \ldots \qquad (9)$$

the NF, Eq. (4) becomes

$$\left( -\sum_{i=0} \varepsilon^i v_i \right) u_\xi \equiv \left( -\sum_{i=0} \varepsilon^i v_i \right) S_1[u] = \sum_{i \geq 0} \varepsilon^i a_i S_{i+2}[u] \qquad (a_0 = 1) \qquad (10)$$

In the single-wave case, all the symmetries are proportional to $S_1$, and all the "potentials" – to $G_1$, for any hierarchy for which the symmetries are generated by a linear *recursion operator*:

$$S_n = c_n S_1 \Rightarrow G_n = c_n G_1 \tag{11}$$

(The coefficients $c_n$ are computed by induction.) As a result, the NF merely updates the dispersion relation for $v$. For the "trivial" boundary condition, $u(\xi \to -\infty) = 0$, this relation is well known [2-4]:

$$v = v_0 - \varepsilon a_1 v_0^2 + \varepsilon^2 a_2 v_0^3 - \ldots \tag{12}$$

The proportionality of all the symmetries to $S_1[u]$ leads to relations among many (a priori independent) differential monomials; all the higher derivatives of $u(\xi)$ can be expressed in terms of $u$ itself, $v_0$ (the zero-order approximation to the wave velocity) and, depending on the PDE involved, some low-order derivatives of $u$. For example, for the Burgers equation, only $u$ is required, whereas for the KdV equation – both $u$ and $u_x$ are needed. This drastically reduces the number of independent terms in the homological equation, making it possible to account for all the terms in that equation, so that no obstacles are encountered.

The following quantities will turn out to be especially useful:

$$R_{mn}[u] = S_m[u]G_n[u] - S_n[u]G_m[u] \tag{13}$$

When $u$ is a *single-wave* zero-order solution, then $R_{mn}[u] = 0$ due to Eq. (11). When $u$ is not a single-wave solution, $R_{mn}[u] \neq 0$. They are the building blocks of the canonical form of obstacles.

In the general case, when $u$ is *not a single-wave*, if $u^{(n)}$ is not allowed to depend explicitly on $t$ and/or $x$, obstacles to integrability may emerge in Eq. (8) if the independent differential monomials in $u^{(n)}$ cannot account for all the monomials in $Z_n[u]$. The obstacles are differential polynomials. Their structure is not unique and depends on the algorithm used for cancellation of terms in attempting to solve Eq. (8). As the obstacles are not symmetries, their incorporation in the NF spoils the integrability of the latter. To avoid obstacles, we allow explicit dependence of $u^{(n)}$ on $t$ and/or $x$. This enables the incorporation of the obstacles in the NIT, leaving the NF intact. In addition, we choose the following form for the terms in the NIT:

$$u^{(n)}(x,t) = u_s^{(n)}[u] + u_r^{(n)}(x,t) \tag{14}$$

In Eq. (14), $u_s^{(n)}[u]$ is the differential polynomial that solves the $n$'th-order homological equation in the case of a single-wave solution for the NF. However, it is not computed at the single-wave solution, but at $u$, the solution of the NF in the general case. $u_r^{(n)}(x,t)$ depends explicitly on $x$ and $t$. Its role is to account for the effect of the obstacles. It is determined by the following equation:

$$\partial_t u_r^{(n)} = \left[ S_2[u], u_r^{(n)} \right] + R^{(n)} \tag{15}$$

$R^{(n)}$ is the obstacle in the homological equation of order $n$, i.e., the sum of all terms that would have been unaccountable for, if $u^{(n)}(x,t)$ were not allowed to depend explicitly on $t$ and $x$. Now, that the NIT accounts for the effect of the obstacles, there is no need to assign them to the NF. The latter remains integrable, as it is constructed solely from symmetries, and, hence, preserves the nature of the unperturbed solution.

As $u_s^{(n)}[u]$ has been computed so as to solve the $n$'th order homological equation in the case of a single-wave solution, it cancels in the general case all the terms in the homological equation except for those that vanish identically for the single-wave solution, but are present now. These are the obstacles. Their generic form is:

$$R^{(n)}[u] = \sum \gamma_{pq}^{(n)} f_{pq}^{(n)}[u, \partial_x] R_{pq}[u] \tag{16}$$

In Eq. (16), $\gamma_{pq}^{(n)}$ are constant coefficients; $f_{pq}^{(n)}[u, \partial_x]$ are operators that ensure that the obstacle has the correct weight [14]. They may include differential polynomials in $u$, as well as operators of differentiation, $\partial_x$, and indefinite integration, $\partial_x^{-1}$, acting on $R_{pq}[u]$.

We have analyzed the perturbed Burgers, KdV and heat-diffusion equations. To show the advantage of canonical obstacles, we study them for the case of a double-wave solution.

The perturbed Burgers equation

$$w_t = 2ww_x + w_{xx} + \varepsilon\left(3\alpha_1 w^2 w_x + 3\alpha_2 ww_{xx} + 3\alpha_3 w_x^2 + \alpha_4 w_{xxx}\right) + \\ \varepsilon^2\left(4\beta_1 u^3 u_x + 6\beta_2 u^2 u_{xx} + 12\beta_3 uu_x^2 + 4\beta_4 uu_{xxx} + 10\beta_5 u_x u_{xx} + \beta_6 u_{xxxx}\right) + O(\varepsilon^3) \quad (17)$$

has the Normal Form

$$u_t = 2uu_x + u_{xx} + \varepsilon\alpha_4\left(3u^2 u_x + 3uu_{xx} + 3u_x^2 + u_{xxx}\right) + \\ \varepsilon^2\beta_6\left(4u^3 u_x + 6u^2 u_{xx} + 12uu_x^2 + 4uu_{xxx} + 10 u_x u_{xx} + u_{xxxx}\right) + O(\varepsilon^3) \quad (18)$$

We first focus on the <u>single-wave solution</u> of the NF, whose form is

$$u(x,t) = \frac{kB\exp(k(x-vt))}{1 + B\exp(k(x-vt))}, \quad v = -k - \varepsilon\alpha_4 k^2 - \varepsilon^2 \beta_6 k^3 + O(\varepsilon^3) \quad (19)$$

The first-order correction, $u_s^{(1)}[u]$, is given by

$$u_s^{(1)}[u] = (\alpha_1 - 2\alpha_2 - \alpha_3 + 2\alpha_4)q u_x - \frac{1}{2}(2\alpha_1 - \alpha_2 + \alpha_3 - 2\alpha_4)u^2 \quad \left(q \equiv \int_{-\infty}^{x} u(x,t)dx\right) \quad (20)$$

and no obstacles emerge, because the only obstacle possible in this order (in the canonical form, see Eq. (13)) vanishes in the case of a single-wave solution:

$$R_{21} = S_2 G_1 - S_1 G_2 = u^2 u_x + uu_{xx} - u_x^2 = 0 \quad (21)$$

We now turn to the <u>double-wave solution</u> of Eq. (18)

$$u(x,t) = \frac{k_1 B_1 \exp(k_1(x-v_1 t)) + k_2 B_2 \exp(k_2(x-v_2 t))}{1 + B_1 \exp(k_1(x-v_1 t)) + B_2 \exp(k_2(x-v_2 t))}, \quad v_i = -k_i - \varepsilon\alpha_4 k_i^2 - \varepsilon^2 \beta_6 k_i^3 + O(\varepsilon^3) \quad (22)$$

The first-order correction, $u^{(1)}$, is constructed according to Eq. (14). $u_s^{(1)}[u]$ is given by Eq. (20), now with $u$ of Eq. (22). The correction term, $u_r^{(1)}$, obeys Eq. (15), which now has the form:

$$\partial_t u_r^{(1)} = 2 u_x u_r^{(1)} + 2 u \partial_x u_r^{(1)} + \partial_x^2 u_r^{(1)} + \gamma_{21} R_{21} \tag{23}$$

where the obstacle, $R_{21}$, does not vanish identically, and

$$\gamma_{21} = 2\alpha_1 - \alpha_2 - 2\alpha_3 + \alpha_4 \tag{24}$$

For $k_1 \cdot k_2 < 0$, the two fronts in the zero-order solution of Eq. (22) are distinct for $t \ll 0$. Near the origin in time they coalesce and form a single front which persists for all $t > 0$ and is centered around the line $x = -(k_1+k_2) \cdot t$ (see Fig. 1a). Thus, we interpret $t \geq 0$ as the interaction region. The asymptotic behavior of the canonical obstacle $R_{21}$ is:

$$R_{21} \to \begin{cases} 0 & t \to -\infty \\ k_1 k_2 u_x & t \to +\infty \end{cases} \tag{25}$$

(see Figs. 1b and 1c). The solution for $u_r^{(1)}$ around the origin has to be found numerically. However, its asymptotic form away from the origin is:

$$u_r^{(1)} = \begin{cases} -\tfrac{1}{2} \gamma_{21} k_1 k_2, & t \gg 0 \\ 0 & t \ll 0 \end{cases} \tag{26}$$

For $k_1 \cdot k_2 > 0$, the limits for $t \gg 0$ and $t \ll 0$ are interchanged.

The $O(\varepsilon^2)$ obstacle is also constructed from canonical terms, which are finite only in the region of interaction, and so will be $u_r^{(2)}$. We do not present the $O(\varepsilon^2)$ results in detail because the term proportional to $q(x,t)$, already in Eq. (20), although permitted by the NF formalism, is unbounded (asymptotically it is linear in $x$), limiting the validity of the approximation to $|t|$ and $|x|$ of $O(1)$. Thus, the perturbed Burgers equation should be viewed primarily as a theoretical tool. It is a simple example for the emergence of a first-order obstacle in a perturbed regular PDE.

The perturbed KdV equation:

$$w_t = 6ww_x + w_{xxx} + \varepsilon\left(30\alpha_1 w^2 w_x + 10\alpha_2 w w_{xxx} + 20\alpha_3 w_x w_{xx} + \alpha_4 w_{5x}\right) +$$
$$+\varepsilon^2\begin{pmatrix} 140\beta_1 w^3 w_x + 70\beta_2 w^2 w_{xxx} + 280\beta_3 w w_x w_{xx} + \\ +14\beta_4 w w_{5x} + 70\beta_5 w_x^3 + 42\beta_6 w_x w_{4x} + 70\beta_7 w_{xx} w_{xxx} + \beta_8 w_{7x} \end{pmatrix} + O(\varepsilon^3) \quad (27)$$

has the Normal Form

$$u_t = 6uu_x + u_{xxx} + \varepsilon\alpha_4\left(30u^2 u_x + 10u u_{xxx} + 20u_x u_{xx} + u_{5x}\right) +$$
$$+\varepsilon^2\beta_8\begin{pmatrix} 140 u^3 u_x + 70 u^2 u_{xxx} + 280 u u_x u_{xx} + \\ +14 u u_{5x} + 70 u_x^3 + 42 u_x u_{4x} + 70 u_{xx} u_{xxx} + u_{7x} \end{pmatrix} + O(\varepsilon^3) \quad (28)$$

The single-wave solution of the NF is

$$u(x,t) = 2k^2 \operatorname{sech}^2\{k(x + vt + x_0)\}, \quad v = 4k^2 - 16\varepsilon\alpha_4 k^4 + 64\varepsilon^2 \beta_8 k^6 + O(\varepsilon^3) \quad (29)$$

and the first-order correction, $u_S^{(1)}$, is given by

$$u_s^{(1)}[u] = \frac{1}{6}(6\lambda - 15\alpha_1 + 10\alpha_2 - 10\alpha_3 + 15\alpha_4)u^2 +$$
$$\frac{1}{6}(6\lambda + 15\alpha_1 - 20\alpha_2 - 10\alpha_3 + 15\alpha_4)q^{(1)}u_x + \lambda u_{xx} \quad \left(q^{(1)} \equiv \int_{-\infty}^{x} u(x',t)dx'\right) \quad (30)$$

where λ is undetermined. (In fact, the coefficient of either of the monomials in Eq. (30) may be equivalently chosen as a free parameter.) No obstacles emerge in the single-wave case, because the only possible canonical obstacle in first order (see Eq. (13)) vanishes:

$$R_{21} = S_2 G_1 - S_1 G_2 = 3u^2 u_x + u u_{xxx} - u_x u_{xx} = 0 \quad (31)$$

The form of the second-order correction in the single-wave case is

$$u_s^{(2)} = b_1 u^3 + b_2 u u_{xx} + b_3 u_x^2 + b_4 u_{4x} + b_5(6u u_x + u_{xxx})q^{(1)} + b_6 u_{xxx} q^{(1)} + b_7 u_{xx} q^{(1)^2} + b_8 u_x q^{(2)} \quad (32)$$

where

$$q^{(2)} \equiv \int_{-\infty}^{x} u^2(x,t)dx \quad (33)$$

The formalism can determine all the coefficients $b_i$, except for two, chosen here to be $b_5$ and $b_8$.

We now turn to the two-soliton solution of the NF, Eq. (28), which may be expressed by the Hirota formula [19]:

$$u(x,t) = 2\partial_x^2 \ln\left\{1 + f_1 + f_2 + \left(\frac{k_1 - k_2}{k_1 + k_2}\right)^2 f_1 f_2\right\} \qquad \left(f_i = \exp\left[2k_i(x + v_i t)\right]\ i = 1,2\right) \qquad (34)$$

where

$$v_i = 4k_i^2 - 16\varepsilon\alpha_4 k_i^4 + 64\varepsilon^2 \beta_8 k_i^6 + O(\varepsilon^3) \qquad (35)$$

Away from the interaction region of the two solitons, Eq. (34) is asymptotically reduced to a superposition of two single-soliton solutions (see Fig. 2a):

$$u(x,t) = 2k_1^2 \operatorname{sech}^2\{k_1(x - v_1 t)\} + 2k_2^2 \operatorname{sech}^2\{k_2(x - v_2 t)\} \qquad (36)$$

The first-order correction, $u^{(1)}$, is again constructed according to Eq. (14), with $u_s^{(1)}[u]$ given by Eq. (30). Before the freedom of choice of $\lambda$ is exploited, an obstacle emerges here as well. With our choice of $u_s^{(1)}[u]$, this obstacle has the canonical form:

$$R^{(1)} = \gamma_{21}^{(1)} R_{21} \qquad \left(\gamma_{21}^{(1)} = \frac{1}{2}(6\lambda + 15\alpha_1 - 10\alpha_3 - 5\alpha_4)\right) \qquad (37)$$

This first-order obstacle is eliminated by choosing $\lambda = -\frac{5}{6}(3\alpha_1 - 2\alpha_3 - \alpha_4)$. (That no obstacle is encountered in the first-order analysis of the perturbed KdV equation is well known [2]. With our algorithm, both the emergence of the canonical obstacle and its elimination are explicit.) Thus, in the case of the perturbed KdV equation, the correction term, $u_r^{(1)}(x,t)$, vanishes.

In second order, using Eq. (14), the structure of the obstacle is found to be:

$$R^{(2)} = \left[\left(\tfrac{1}{2}(3b_5 - b_8) + c_1\right)u + \left(\tfrac{1}{2}(3b_5 - b_8) + c_2\right)\partial_x^2 + \left(\tfrac{1}{2}(3b_5 - b_8) + c_3\right)u_x \partial_x^{-1}\right]R_{21} + \\ \left(\tfrac{1}{2}(3b_5 + b_8) + c_4\right)R_{31} \qquad (38)$$

$c_k$ are known combinations of the $\alpha$ and $\beta$ coefficients of Eq. (27). (An appropriate choice of $b_5$ and $b_8$ may be used to eliminate $R_{31}$ and $u_x \partial_x^{-1} R_{21}$.) As the obstacle $R^{(2)}$ is constructed from canonical obstacles, it vanishes identically in the single-soliton case. In the general case, $R^{(2)}$ does not vanish identically and is accounted for by $u_r^{(2)}(x,t)$, which obeys the following equation:

$$\partial_t u_r^{(2)}(x,t) = 6 u_x u_r^{(2)} + 6 u \partial_x u_r^{(2)} + \partial_x^3 u_r^{(2)} + R^{(2)} \qquad (39)$$

Near the origin, the two solitons interact; $u_r^{(2)}(x,t)$ has to be found numerically. Away from the interaction region ($|t| \to \infty$), the obstacle vanishes exponentially (see Fig. 2b). Thus, its effect is not felt, and the second-order correction preserves the two-wave nature of the solution.

Our last example is the heat-diffusion equation with a first-order perturbation:

$$w_t = w_{xx} + \varepsilon \left( \alpha_1 w^2 w_x + \alpha_2 w w_{xx} + \alpha_3 w_x^2 + \alpha_4 w_{xxx} \right) + O(\varepsilon^2) \qquad (40)$$

For the sake of brevity, we only mention here the most important findings. Again, there are no obstacles in the single-wave case, whereas an obstacle does emerge in the double-wave case

$$u(x,t) = A_1 \exp[k_1(x - v_1 t)] + A_2 \exp[k_2(x - v_2 t)]; \qquad v_i = -k_i - \varepsilon \alpha_4 k_i^2 - O(\varepsilon^2) \qquad (41)$$

Luckily, the freedom in the expansion enables us to obtain an expression for $u_s^{(1)}[u]$, for which the first-order obstacle vanishes in the anti-symmetric case, $k_2 = -k_1$. If the zero-order approximation is not anti-symmetric, one can perform a Galilean transformation to a moving frame in which the wave numbers obtain anti-symmetric values, so that the effect of the obstacle can be eliminated.

Vanishing of obstacles for particular solutions is not specific to the two-wave solution of Eq. (40). In the case of the perturbed Burgers equation, the canonical obstacles, $R_{mn}$, vanish in the anti-symmetric case of the two-front solution, Eq. (22), when $(m + n)$ is even. These results suggest a possible way to eliminate some or all of the obstacles (depending on the problem considered):

*i)* Transform the given problem to an anti-symmetric one using a Galilean transformation;

*ii)* Construct the NIT so that the resulting obstacle vanishes for the anti-symmetric solution.

In summary, obstacles to integrability can be accounted for by allowing the higher orders in the NIT to depend on $t$ and $x$ explicitly. The NF remains integrable, as it is constructed from symmetries only. The algorithm of Eq. (14) for the construction of the NIT generates "canonical" obstacles, which vanish explicitly in the single-wave case. For multi-wave solutions of the NF, the canonical obstacles are amenable to simple predictions. They are sizeable only in the region of interaction among the waves, and fall off exponentially towards zero away from it. Hence, they affect the NIT only in this region. For instance, for localized solutions (e.g., multi-solitons of the perturbed KdV equation) the obstacles do not affect the NIT for $|t|\rightarrow\infty$. For systems whose solutions are wave fronts (e.g., the perturbed Burgers equation) the interaction region may extend over an infinite range in $t$. Still, the obstacles vanish exponentially away from the interaction region. As the examples studied here indicate, the effect of the obstacles on the asymptotic behavior of $u^{(n)}$, the correction terms in the NIT, can be obtained in closed form.

In the usual analysis [5-12], the (non-canonical) structure of obstacles is not unique; it is difficult to quantify their properties and their effect on the solution, or to show in a simple way that they do not emerge in the case of a single-wave solution. The obstacles may be sizeable away from the interaction region, and affect the higher-order computation in a complicated manner. For instance, one may be led to the conclusion that the individual solitons undergo inelastic scattering [7, 10, 15-18]. Clearly, in a given order of the expansion, different ways of solving the perturbed PDE provide approximate solutions of the same numerical quality. However, the physical interpretation of the results depends on the details of the expansion algorithm.

Figure captions

Fig. 1  Two-front solution of Burgers equation (se Eq. (22)); $B_1 = B_2 = 1$, $k_1 = 0.5$, $k_2 = -0.5$
  a – $u(x,t)$, b – $u_x(x,t)$, c – Canonical obstacle $R_{21}$ (Eq.(13))

Fig. 2  Two-soliton solution of KdV equation (see Eq. (34)): $k_1=0.5$, $k_2=-0.75$
  a – $u(x,t)$, b – Canonical obstacle $R_{21}$ (Eq.(13))

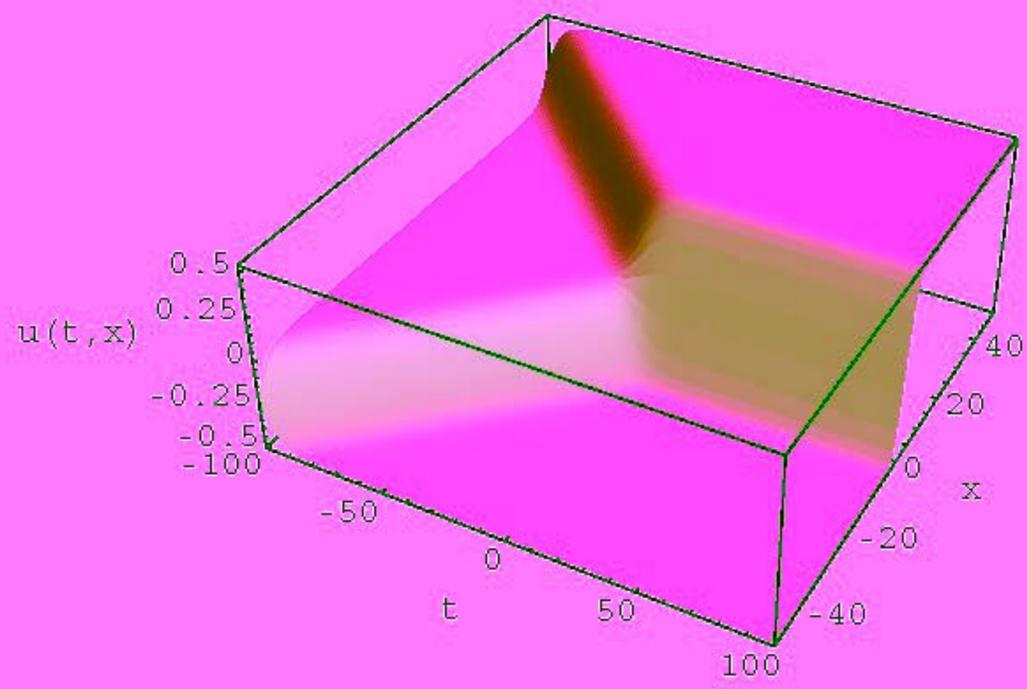

Fig. 1a

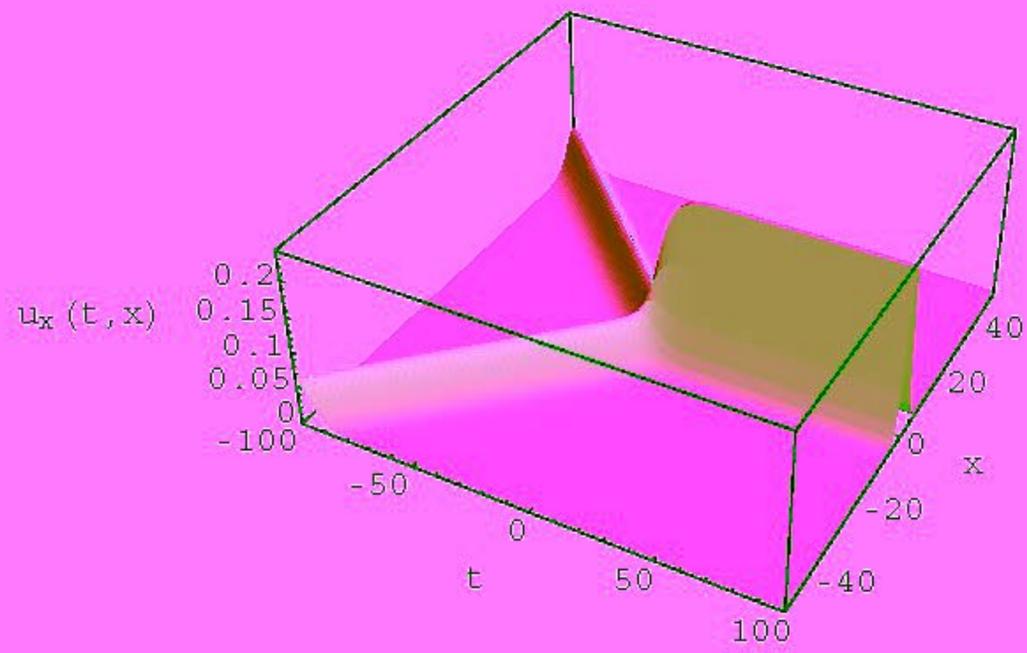

Fig. 1b

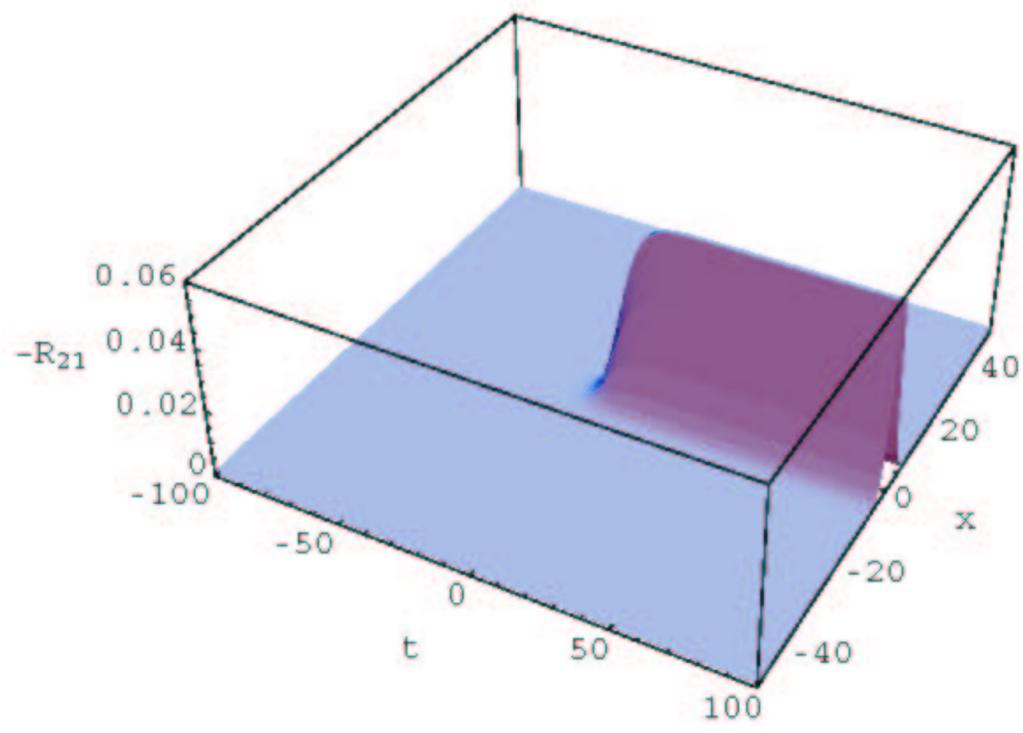

Fig. 1c

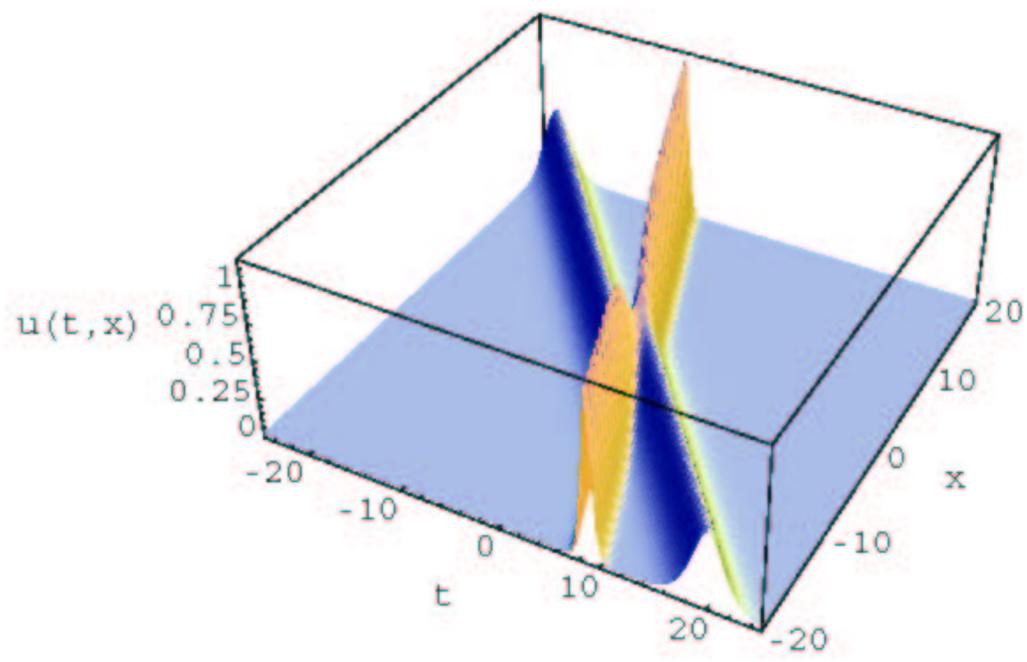

Fig. 2a

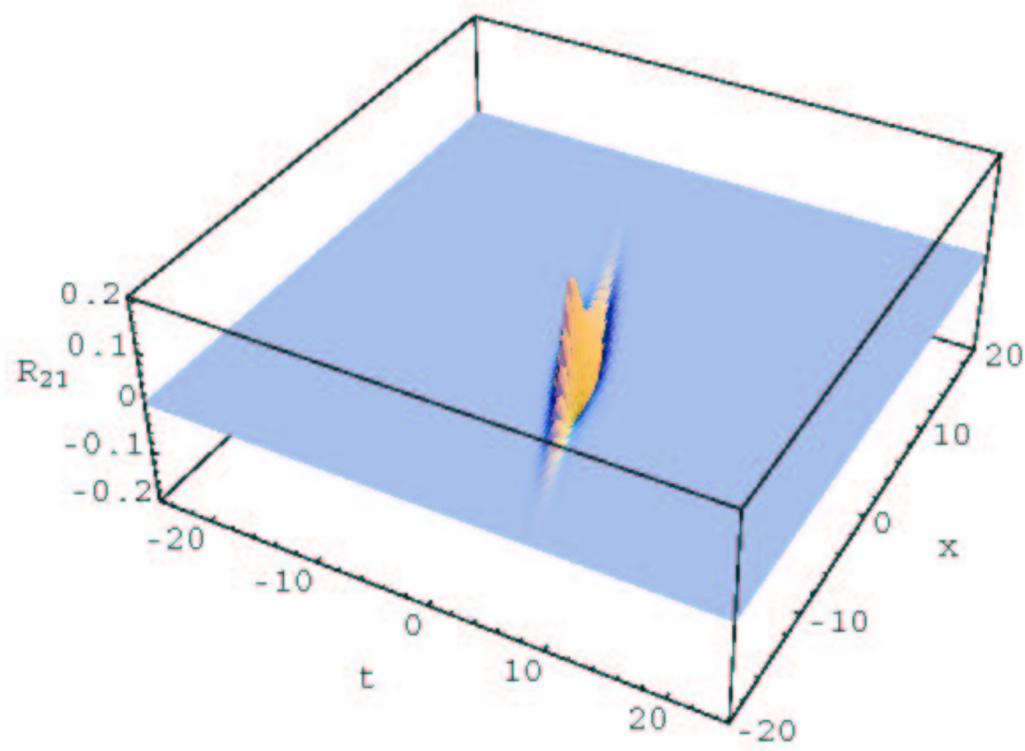

Fig. 2b